\documentclass[aps,showpacs,twocolumn]{revtex4}
\usepackage{amsmath}
\usepackage{graphicx}

\begin{document}

\title{Structure of the Particle-Hole Amplitudes in No-core Shell Model
Wave Functions}

\author{A.C. Hayes$^1$, A.A. Kwiatkowski$^2$}
\affiliation{$^1$Theoretical Division, Los Alamos National Laboratory,
Los Alamos, New Mexico 87545 \\
$^2$National Superconducting Cyclotron Laboratory, Michigan State University, East Lansing, Michigan 48824 \\
}

\begin{abstract}
We study the structure of the no-core shell model wave functions for $^6$Li and
$^{12}$C  by investigating the ground state and first excited state electron scattering
charge form factors.
In both nuclei, large particle-hole ($ph$) amplitudes in the wave functions appear with
the opposite sign to that needed to reproduce the shape of the $(e,e')$ form factors,
the charge radii, and the B(E2) values for the lowest two states.
The difference in sign appears to arise mainly from the monopole 
$\Delta\hbar\omega=2$  matrix elements of the kinetic and potential energy (T+V)
that transform under the harmonic oscillator SU(3) symmetries as $(\lambda,\mu)=(2,0)$.
These are difficult to determine self-consistently, but
they have a strong effect on the structure of the low-lying states
and on the giant monopole and quadrupole resonances.
The Lee-Suzuki transformation, used to account for the restricted nature of the space in terms of an effective interaction, introduces large higher-order $\Delta\hbar\omega=n, n>$2, $ph$
amplitudes in the wave functions.  The latter $ph$ excitations
 aggravate the disagreement between the experimental and predicted $(e,e')$ form factors
with increasing  model spaces, especially at high momentum transfers.
For sufficiently large model spaces the situation begins to 
 resolve itself for $^6$Li, but the
convergence is  slow. A prescription to constrain the $ph$ excitations would likely
accelerate convergence of the calculations.
\end{abstract}
\pacs{21.60.-n,21.60.Cs,21.60.De,24.30.Cz,25.30.Bf,25.30.Dh  }
\maketitle

\section{Introduction}
The  {\it ab initio} no-core shell  model (NCSM) permits  calculations of wave functions
in very large model-space sizes for nuclei at the beginning of the p-shell.
For $^6$Li calculations up to 16$\hbar \omega$ have been achieved \cite{navLi}.
Among the  successes of  the model
is its predicted energy spectra of light nuclei \cite{navLi,navEnergy1,navEnergy2}.
Towards the end of the p-shell a $10\hbar\omega$ basis calculation has been achieved for A=11 \cite{forssen}. In mass 12 the model provides a reasonable description  of the low-momentum component
 of the vector and axial currents involved in the
electro-weak transitions to  the ground state
on $^{12}$N when a three-body interaction is included \cite{navC}.
In all of these calculations, a Lee-Suzuki \cite{lee-suzuki} transformation of the nucleon-nucleon interaction is used to account for the restricted nature of the space in terms of an effective interaction. 
Group theoretical analyses \cite{dytrych} of the no-core shell model wave functions 
have shown that the predicted eigenstates of $^{12}$C and $^{16}$O have very large overlaps
with a small sub-space of the full model space, with the sub-space being defined by the most deformed symplectic basis states.
The purpose of this paper is to examine
the structure of the multi-$\hbar\omega$ terms in the wave functions
for the low-lying
states at the beginning and end of the $p$-shell in more detail. For this we
compare
NCSM predictions with measured elastic and inelastic $(e,e')$ charge form factors in $^6$Li and $^{12}$C.

The shape of the electron scattering form factors provides a direct probe
of the magnitude and structure of the higher shell components in the wave functions.
The charge form factors have the additional advantage that two-body meson-exchange
currents do not contribute significantly to the form factors
below about 2 fm$^{-1}$ \cite{carlson}.
In all $(e,e')$ calculations presented here we use a bare one-body operator;
as discussed below, the introduction of an effective operator (to compensate for the truncated model space) does
not significantly affect our conclusions.

 Our first main finding is that the $ph$ amplitudes that contribute significantly to electron scattering appear with the opposite sign to that needed to replicate the experimental form factors, elastic and inelastic, and the charge radii.  The 2$\hbar\omega$ contributions to the inelastic form factor change sign, in agreement with experiment, for sufficiently large model spaces for $^6$Li; however, higher-order terms do not within the model spaces we examined.  Second we show that the symplectic $(\lambda,\mu)$, $\Delta\hbar\omega$ = 2 $ph$ amplitudes in the wave functions
are sensitive functions of the oscillator parameter.

\section{The Elastic C0 Form Factors}
The ground state C0 form factor is the Fourier transform of the charge density, and contributions from two-body charge operators
and/or relativistic corrections are negligible
for momenta up to about 2 $fm^{-1}$ \cite{carlson}.
In a harmonic oscillator (HO) basis, the  $0\hbar\omega$ $p$-shell charge form factor
 is given by
\begin{equation}
F_{0p-0p }(q^2)= \sqrt{3} (1-2/3y) exp(-y)
\end{equation}
where $y=(bq/2)^2$ and $b$ is the oscillator parameter.

When additional shells are added to the model space the new contributions to the form factor fall into two main classes.
The first of these are the {\it in-shell} contributions (e.g. $1s0d-1s0d, 1p0f-1p0f$, etc.)
 determined by the occupation numbers for the higher shells,
 and the second are from {\it cross-shell} $ph$ excitations (e.g. $0s-1s, 0p-1p$, $0s-2s$, etc.).
At  low $q$ the form factor is determined by the charge radius

\begin{equation}
F(q^2) =1 - \frac{<r^2>\;q^2}{6} + O(q^4)
\end{equation}
and the  higher in-shell contributions can be shown {\it always} to add constructively to the charge radius.
The cross-shell excitations (which for HO wave functions contribute to the charge radius only for
$\Delta\hbar\omega = 2 \; ph$ excitations across two shells) can add constructively or destructively.
In a $2\hbar\omega$ calculation for p-shell nuclei there are two possible cross-shell contributions, namely,

\begin{equation}
F_{0s-1s}(q^2) = \sqrt{2/3} y\; exp(-y)
\end{equation}

$$
F_{0p-1p}(q^2) =  \sqrt{10/3} y (1-2/5 \;y)\; exp(-y)
$$
Both of these transform under SU(3) as $(\lambda,\mu)=(2,0)$ and represent the $2\hbar\omega$ symplectic
contributions to the form factors.
When these and/or higher shell $\Delta\hbar\omega=2$ $ph$ excitations appear with a sign so as to enhance the predicted charge radius,
they  pull in the charge form factor  in momentum space.

\subsection{C0 form factor for $^6$Li}
Both elastic and inelastic scattering from $^6$Li
have been studied extensively \cite{his1,his2,his3,stanford,saskatoon,mainz}.
Our calculations use the CD Bonn nucleon-nucleon interaction \cite{cdbonn}.
The predicted ground state energy of $^6$Li
is least sensitive to the choice
of the oscillator parameter $b$ over increasing
basis size for  $b$ = 1.79 fm ($\hbar\omega$ = 13 MeV) \cite{navLi}, which is the value we use in the present (e,e') calculations.
We note that, because of the increased computational difficulties in calculating the
transition density matrix elements needed for the (e,e') form factors,
 the present calculations are restricted to a maximum model space
of $14\hbar\omega$, to be compared with the 16$\hbar\omega$ model spaces used to calculate the
 energy spectra for $^6$Li.

Figure 1 compares the experimental data for the elastic form factor for $^6$Li with the model predictions.
For model spaces up to 10$\hbar\omega$, the predicted form factor moves out further in momentum space as the
basis is increased.
For the largest model space examined, $14\hbar\omega$, the tend begins to reverse.
\begin{figure}[h]
\includegraphics[width=2.8in,angle=-90]{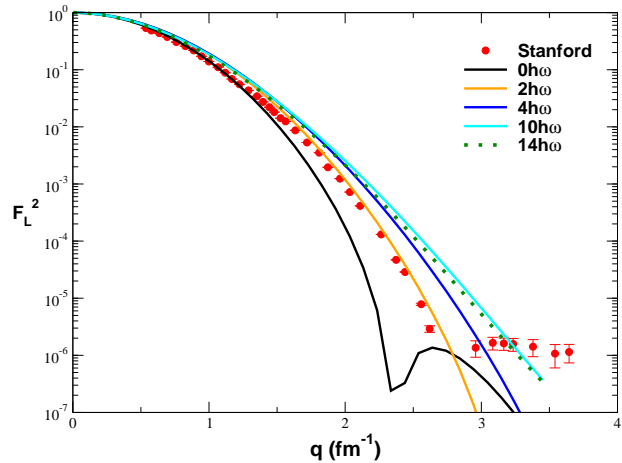}
\caption{(color online)The elastic C0 form factor
for the ground state of $^6$Li.
For all but the largest model space,
the form factor moves further out in $q$ as the model space is increased.  
Experimental data are taken from \cite{stanford}}.

\end{figure}
In coordinate space
the predicted charge density (Figure 2) is enhanced in the interior, with little change to the tail as higher shells are added.
These trends reflect the structure and sign of  $ph$ excitations introduced
as the model space increases.
\begin{figure}[h]
\includegraphics[width=2.8in,angle=-90]{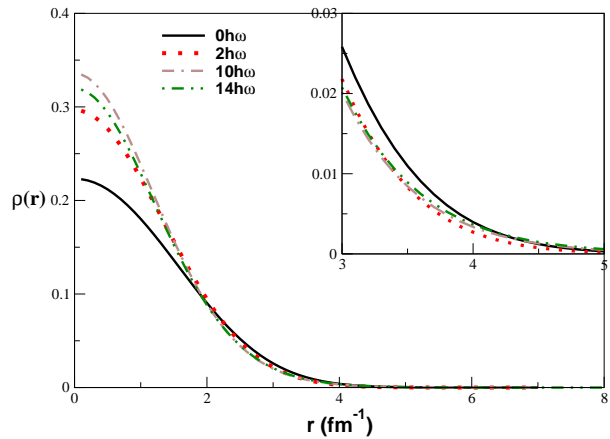}
\caption{(color online) The ground state charge density of $^6$Li. As the model space increases the density
is enhanced in the interior, with minor changes in the tail region.
For sufficiently large model spaces the strength in the interior starts to become suppressed
with a corresponding build up in the tail region.
}
\end{figure}
 There are two issues with
 the structure of the predicted ground state $ph$ excitations.
 First,  the $\Delta\hbar\omega = 2 $ $ph$ excitation for all  shells included  add destructively to the
ground state charge radius. These suggest that the sign of the important symplectic excitations in the wave functions
may be problematic. 
 In momentum-space these  excitations pull the form factor out in $q$.
\begin{figure}[t]
\includegraphics[width=2.8in,angle=-90]{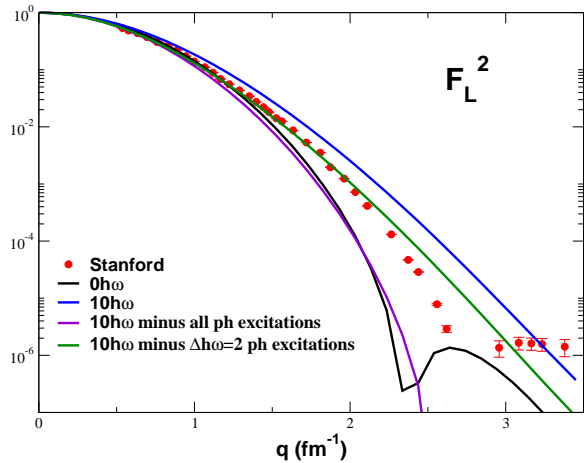}
\caption{color online) The ground state C0 form factor for $^6$Li. The figure displays the effect of particle-hole excitations on the predicted form factor by arbitrarily setting the one-body density matrix elements to zero.  Experimental data are taken from \cite{stanford}.
}
\end{figure}
\begin{figure}
\includegraphics[width=2.8in,angle=-90]{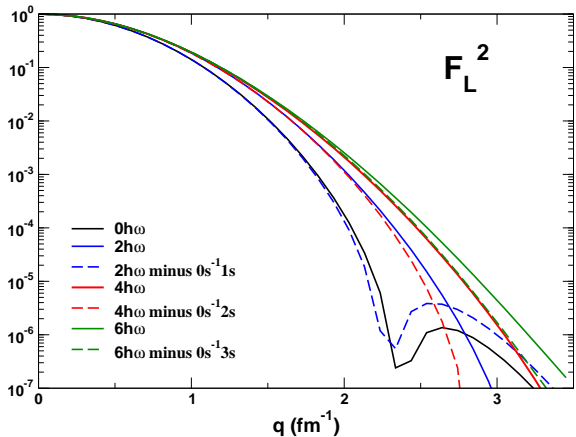}
\caption{(color online) The ground state C0 form factor for $^6$Li. The figure displays the effect of $0s\rightarrow ns$ $ph$ excitations on the predicted form factor. The dashed
curves show the effect of setting the one-body density matrix elements for these excitation to zero (arbitrarily).
 The $0s\rightarrow ns$ excitations
are sizable and pull the form factor out further at higher momentum transfers.}
\end{figure}
Second, there are large $0s\rightarrow ns$ excitations which  pull the form factor out further at higher $q$.
These effects are shown in Figures 3 and 4, where the relevant $ph$ one-body density matrix elements have been arbitrarily set to zero for the purposes
of displaying their effect on the shape of the predicted form factor.

In Table 1 we  show the contributions to the charge radius from in-shell versus cross-shell excitations, and the destructive interference
from the $ph$ excitations is  larger than the constructive interference from the higher in-shell excitations.
This
suggests that the predicted sign of the $\Delta\hbar\omega =2 $ $ ph $ excitations that transform under SU(3) as $(\lambda,\mu$)=(2,0) inhibits the convergence of the calculations.

\begin{table}
\begin{tabular}{|c |c  c c c  c c|}
\hline
\multicolumn{7}{|c|}{Point charge radius  for $^6$Li in units of fm}\\
\hline
Model space &$0\hbar\omega$ & $2\hbar\omega$ & 4$\hbar\omega$ & 10$\hbar\omega$&14$\hbar\omega$& Expt.\\
\hline
full model space& 2.23 & 2.08 & 2.1 & 2.14 & 2.21&2.38$\pm$ 0.1\\
$ph$ contributions omitted &2.23 & 2.26 & 2.31&2.38&2.41&\\
\hline
\end{tabular}
\caption{}
\end{table}

\subsection{C0 form factor for $^{12}$C}
\begin{figure}[h]
\includegraphics[width=2.8in,angle=-90]{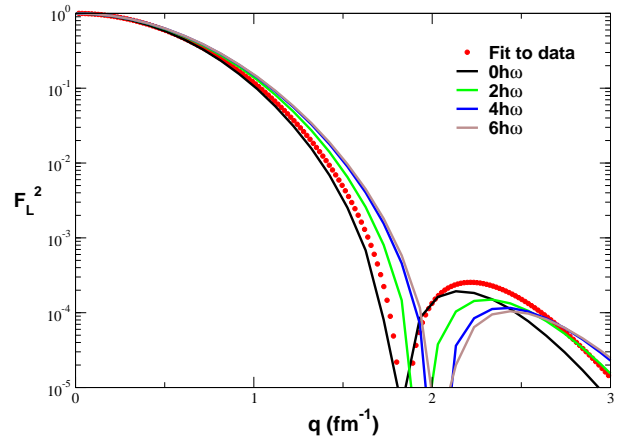}
\caption{(color online) The elastic C0 form factor
for the ground state of $^{12}$C.
As the model space increases the form factor moves further out in $q$, and the
dominant  higher shell terms add destructively to the predicted charge radius.
Experimental data were taken from \cite{exp1, exp2, exp3,exp4}.}
\label{12c-elas}
\end{figure}

\begin{figure}[t]
\includegraphics[width=2.8in,angle=-90]{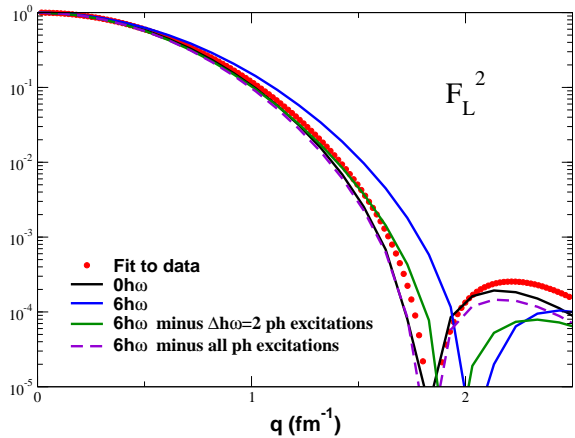}
\caption{(color online) The elastic C0 form factor
for the ground state of $^{12}$C.
The 6$\hbar\omega$ calculations displaying the effect of the removal of the $ph$ excitations were obtained
by arbitrarily setting the corresponding one-body density matrix elements to zero.
}
\label{12c-6hw}
\end{figure}

The trends seen for the elastic scattering form factor for $^6$Li are also seen in the case of $^{12}$C.
Again the calculations use the CD Bonn interaction \cite{cdbonn}.
The $\Delta\hbar\omega = 2$ $ph$ excitations add destructively to the charge radius
 and pull the elastic C0 form factor out in momentum space, Figure \ref{12c-elas}.
We note that the charge radius for $^{12}$C (Table \ref{12c-chRad}) is over-predicted,
which in part reflects the chosen oscillator parameter, $b=1.663$ fm ($\hbar\omega = 15 $MeV),  which minimizes the ground state energy.  We will discuss the choice and effect of $b$ in section \ref{b-choice}.
As in the case of $^6$Li, $ph$ excitations across more than two shells are large, and they act so as to move the
form  factor out further in $q$. For $^{12}$C these include
both the $0s\rightarrow ns$ and $0p\rightarrow np$ excitations.
\begin{table}[h]
\begin{tabular}{|c |c  c c c  c |}
\hline
\multicolumn{6}{|c|}{Point charge radius for $^{12}$C in units of fm}\\
\hline
Model space &$0\hbar\omega$ & $2\hbar\omega$ & 4$\hbar\omega$ & 6$\hbar\omega$& Expt.\\
\hline
full model space& 3.18 & 2.99 & 2.95 & 2.95 & 2.32$\pm$ 0.022\\
$ph$ contributions omitted& 3.18&3.20&3.25 & 3.24&\\
\hline
\end{tabular}
\caption{}
\label{12c-chRad}
\end{table}
Figure \ref{12c-6hw} displays the effect of arbitrarily setting the $ph$ contributions to the form factor to zero.

\section{C2 Form Factors}
The C2 form factor is determined by the transition charge density; there is no significant contribution from two-body meson exchange currents below $q \approx$ 2 fm$^{-1}$\cite{joe}.
 We examine the
longitudinal form factor for scattering to the 2.186 MeV
($3^+$ T~=~0) state in $^6$Li and the 4.44 MeV ($2^+$ T~=~0) state in $^{12}$C.
  Data for the $^6$Li C2 transition  have been measured to $q \approx$ 3.5 fm$^{-1}$ \cite{stanford,saskatoon,mainz}.
Extensive data are also available for the 4.44 MeV (2$^+$) state in $^{12}$C \cite{exp1,exp2,exp3,exp4}.

The most significant contributions to the C2 form factors for
$p$-shell nuclei in a $(0+2)\hbar\omega$ calculation are transitions
within the $p$-shell ($0p\rightarrow 0p$) and $ph$ excitations  across two shells that correspond to the excitation of the GQR. The latter transform under SU(3) as $(\lambda,\mu)L= (2,0)2$.
  For HO wave functions the $0p\rightarrow 0p$ and GQR form factors are \cite{millener}
\begin{equation}
F_{p-p}(q^2) = -\sqrt{8/15}\; y\;\; exp(-y) \end{equation}
$$
F_{GQR}(q^2) = \sqrt{24/15}\; y\;\; (1 - 1/3 y) exp(-y),
$$
and as before $y = (bq/2)^2$.  If a small admixture of the GQR is added to
the $0\hbar\omega$ state so as to enhance the B(E2),
the form factor is  suppressed at high $q$.
For larger model spaces higher powers of $y$ are introduced.

The shape of the  predicted F$_L$ is often displayed in terms of  the C2 matrix element \cite{millener}.
 In general, the C$\lambda$ matrix element is defined in terms of the form factor $F_\lambda$ as,
\begin{equation}B(C\lambda) = f^{-2}\frac{Z^2}{4\pi}\left(\frac{(2\lambda +1)!!}{q^\lambda}\right)^2 F^2_\lambda
\end{equation}
and
\begin{equation}C2(q)\equiv B(C2)^{1/2} = A + By + Cy^2 + ...
\label{C2}
\end{equation}
where $f = f_{SN}f_{c.m.}exp(-y)$, $f_{SN}$ is the single-nucleon charge form factor \cite{Simon} and the center of mass correction is $f_{c.m.} = exp(y/A)$ \cite{tassie}.
 For a HO basis, the number of terms appearing in the polynomial (\ref{C2})
is determined by the number of shells included in the calculation.
 For $p$-shell nuclei the experimental C2($q$) matrix elements for low-lying states generally decreases with increasing $q^2$, i.e., the coefficient ratio $A/B < 0$, where $A > 0$.

\subsection{C2 Form Factor in $^{12}$C}
\begin{figure}[t]
\includegraphics[width=2.8in,angle=-90]{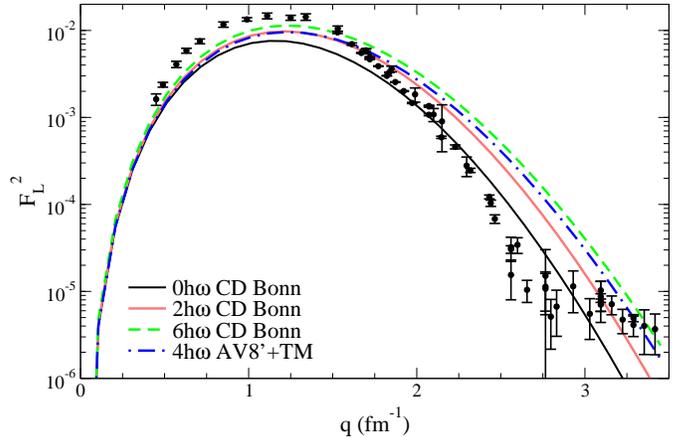}
\caption{(color online) The longitudinal electron scattering form factor
for the 4.44 MeV ($2^+$ T=0) state in $^{12}$C.
As the model space increases the form factor moves further out in $q$,
in contradiction with experimental and theoretical expectations.}
\label{12c-inel}
\end{figure}

Our calculations for $^{12}$C include model spaces up to 6$\hbar\omega$.
We use the CD Bonn \cite{cdbonn} and the AV8' \cite{AV8} nucleon-nucleon interactions,
as well as the AV8' plus the Tucson-Melbourne TM'(99) 3-body \cite{TM} interactions.
The oscillator parameter was taken to be $b = 1.663$ fm $\;\; (\hbar\omega = 15$ MeV).

\begin{figure}
\includegraphics[width=2.8in,angle=-90]{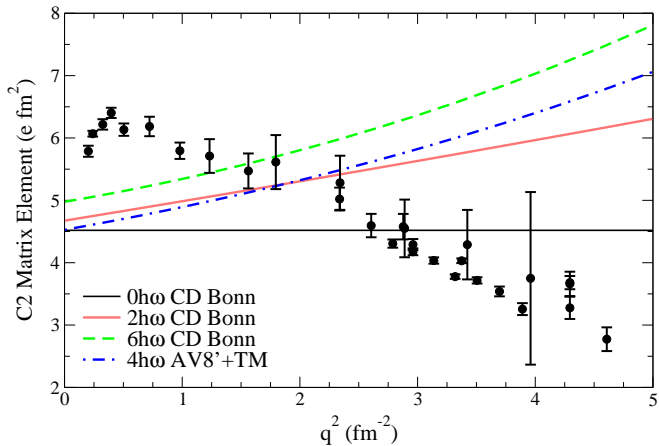}
\caption{(color online) The C2 matrix element
for the 4.44 MeV ($2^+$ T=0) state in $^{12}$C.  The effect of using a three-body interaction (AV8'+TM) is small.  Experimental data were taken from \cite{exp1, exp2, exp3, exp4}}
\label{12c-C2}
\end{figure}
Figure \ref{12c-inel} shows a comparison between the measured and predicted form factors for increasing sizes of the shell model space.  At low momentum transfers the calculations under-predict the form factor.
Above $q \approx 1.5$ fm$^{-1}$ the calculations over-predict the form factor,
and this over-prediction
becomes increasingly worse as the size of the model space is increased.
As the model space is increased beyond $0\hbar\omega$, the
$0p \rightarrow 1p$ excitations
add destructively at  low $q$ (and thus destructively to the
the B(E2) value) and constructively at high $q$, moving the form factor out.
The predicted form factor is enhanced slightly at small $q$ relative to the $0\hbar\omega$ calculation, and we 
 largely attribute this to the
$0s\rightarrow 1d$ excitations, which appear with the correct sign.
We note this differs from the elastic C0 form factors
where excitations from the $0s$ shell appear with the same sign as excitations
from the $0p$ shell.  It should be noted that
the model spaces examined here are restricted to $6\hbar\omega$
and that much larger spaces may well show very different trends.

Figure \ref{12c-C2} displays C2($q$), which we extracted from the measured form factor
using b=1.7 fm.
The experimental C2($q$) matrix element steadily drops with increasing $q^2$.
Our  multi-$\hbar\omega$ calculations predict C2($q$) to have the opposite slope,
in large part because of the sign of the $(0p)^{-1}(1p)$ excitations in the $0^+$ and $2^+$ wave functions.

\begin{figure}
\includegraphics[width=2.8in]{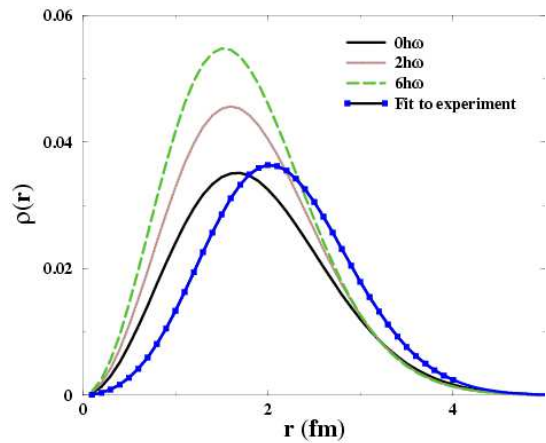}
\caption{(color online) The transition charge density
for the 4.44 MeV ($2^+$ T=0) state in $^{12}$C.
As the model space is increased the transition density moves towards smaller $r$.}
\label{rho-12c}
\end{figure}

Figure \ref{rho-12c} displays the corresponding transition charge density $\rho$(r)
for the $0^+\rightarrow2^+$ transition.  The experimentally determined $\rho(r)$ peaks at about 2 fm, while  the $\rho(r)$ predicted by the NCSM peaks at about 1.5 fm.  As the model space is increased, the peak moves towards smaller $r$.

\subsection{The C2 form factor for $^6$Li}
\begin{figure}[b]
    \includegraphics[angle=270, width=3.75 in]{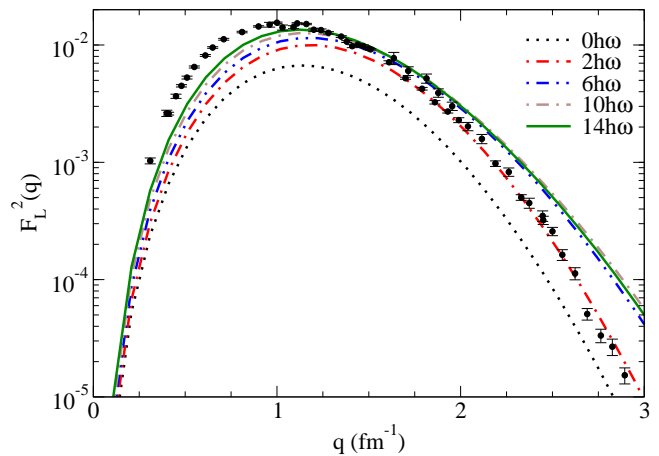}
    \caption{(Color online)  The charge form factor for the $3^+$ 2.186 MeV state using {\it ab initio} NCSM CD Bonn wave functions, where $b$ = 1.79 fm.
 At this value of $b$ ($\hbar\omega =$ 13 MeV), the ground state energy converges fastest over basis size.
 Empirical data are taken from Stanford, Saskatoon, and Mainz \cite{stanford,saskatoon,mainz}. }
    \label{ff13}
\end{figure}

Figure \ref{ff13} displays a comparison of the measured and predicted form factors for increasing basis size for \mbox{$b = 1.79$ fm}.
These calculations use the CD Bonn nucleon-nucleon interaction \cite{cdbonn}.
\begin{figure}[t]
    \includegraphics[angle=270, width=3.75 in]{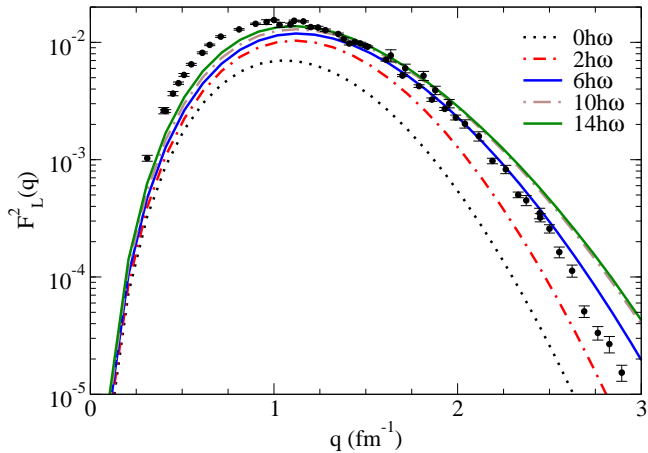}
    \caption{(Color online) The charge form factor is calculated with $b$ = 1.94 fm ($\hbar\omega = $ 11 MeV)
for comparison to figure \ref{ff13}.  The  predictions are suppressed at low $q$ and enhanced at high $q$ for all size model spaces. Experimental data are taken from \cite{stanford,saskatoon,mainz}}
    \label{ff11}
\end{figure}
 At low momentum transfers, $q <$ 1.0 fm$^{-1}$, the calculations  under-predict the form factor.
 At  $q >$ 1.5 fm$^{-1}$, the larger model spaces over-predict the magnitude of the
form factor, and this over-prediction increases with the model space size.
The general trends seen with the sign of higher shell contributions
in the predicted C2 form factors
are very similar to those seen for $^{12}$C. However, our ability to go to considerably higher
shells in the case of $^6$Li allows us to explore these trends in more detail.
The peak of the predicted form factor occurs at higher $q$ than experiment; the biggest
shift of the predicted form factor away from the observed peak occurs between
the 0$\hbar \omega$ and the 2$\hbar \omega$ model space, and a
shift in the peak position to even higher $q$ continues until 8$\hbar \omega$.
Above $8\hbar\omega$ the peak position of the form factor begins to improve. Above the peak, at momenta
$q >$ 1.3 fm$^{-1}$, the additional contributions from higher shell continue to enhance the form factor.
For low momentum transfers (below the peak of the form factor) we see a slow convergence to a magnitude
lower than experiment. This low momentum trend is consistent with the trend of predicted B(E2) values, as summarized in Table \ref{12c-be2}.

\begin{table}
\begin{tabular}{|c |c c c c c c c c |}
\hline
\multicolumn{9}{|c|}{B(E2) Values for $^6$Li in units of e$^2$fm$^4$}\\
\hline
$b$ (fm)  &$0\hbar\omega$ & $2\hbar\omega$ & 4$\hbar\omega$ & $6\hbar\omega$ & $8\hbar\omega$ &10$\hbar\omega$& 14$\hbar\omega$&Expt.\\
\hline
1.94 & 6.84 & 8.14 & 8.93 & 9.89 & 10.73 & 11.63& 13.4&21.8(4.8)\\
1.79 & 4.91 & 6.25 & 7.03 & 8.16 & 9.14& 10.22&12.2&\\
\hline
\end{tabular}
\caption{}
\label{12c-be2}
\end{table}

 We also examined the form factor for a set of calculations with a different oscillator parameter, namely, $b$ = 1.94 fm ($\hbar\omega$ = 11 MeV), Figure \ref{ff11}.  
 The \mbox{$b$ = 1.94 fm} form factors display similar qualitative behavior as the $b$ = 1.79 fm calculations.
 The peak of the former occurs at higher $q$ than experiment and  continues to shift outward until about
10$\hbar \omega$. For large model spaces the situation starts to improve.
 Above the peak of the form factor the higher shell contributions
move the form factor further out in $q$ with increasing model space.

The enhancements for the form factors at large $q$ are determined by the
sign and magnitude of the higher shell contributions in the wave functions.
As in the case of $^{12}$C, this is most striking in the case of the  2$\hbar \omega$ configurations,
where the $0p\rightarrow 1p0f$ $ph$ excitations add destructively at low $q$ and constructively at high $q$.
The slow convergence of the B(E2) to a value smaller than experiment is due in large part
 to the fact that these $ph$ excitations add destructively to the matrix element.

\begin{figure}[t]
    \includegraphics[ width=2.5 in]{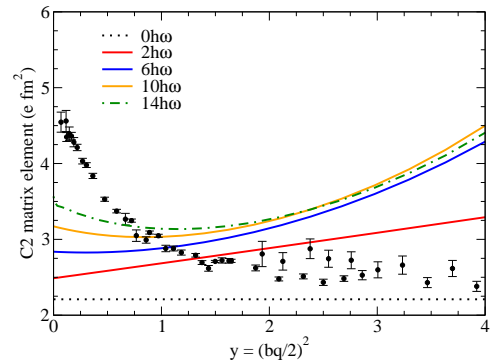}
    \caption{(Color online) The C2 matrix element with $b$ = 1.79 fm.  To calculate the empirical C2 matrix element (data from \cite{stanford,saskatoon,mainz}), we used $b$ = 1.70 fm.}
    \label{c13}
\end{figure}

\begin{figure}[t]
    \includegraphics[ width=2.5 in]{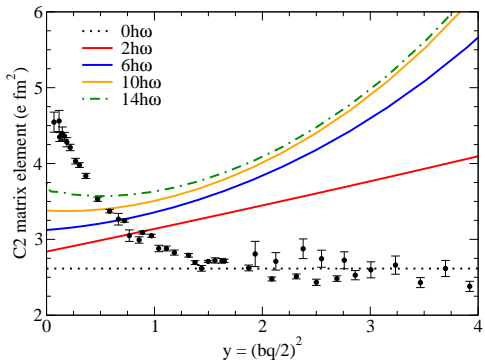}
    \caption{(Color online) The C2 matrix element with $b$ = 1.94 fm.  Experimental data are taken from \cite{stanford,saskatoon,mainz} and a C2 matrix element
extracted using $b$ = 1.70 fm.}  
    \label{c11}
\end{figure}

A more detailed understanding of the convergence of the lower momentum terms in the form factor
with increasing model space can be seen by examining the C2 matrix element.
To obtain an  experimental C2 matrix element, we chose an oscillator parameter $b$ = 1.70 fm, which is close
to the value necessary to give the measured rms charge radius.
The C2 calculated  matrix elements are  displayed in Figures \ref{c13} and \ref{c11}.
 The experimental C2 matrix element decreases with increasing momentum transfer in contrast to the predictions
of the model.
\vspace*{2cm}

\begin{figure}[h]
    \includegraphics[ width=2.5 in]{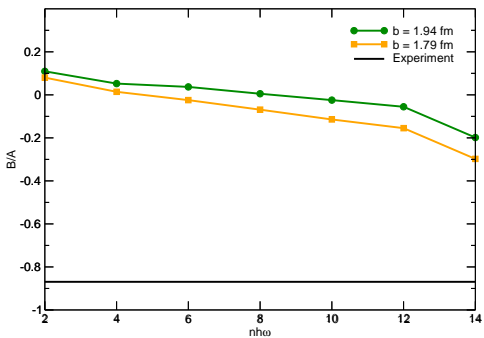}
    \caption{(Color online)  We fitted the C2 matrix element to the polynomial $A +By + Cy^2  \ldots$.  The ratio of $B/A$ is graphed along the $y$-axis.  Since $A$ is always positive, the change in sign is exclusively in $B$.  Experimental data was taken from \cite{stanford,saskatoon,mainz} and  a $B/A$ ratio was extracted using $b$ = 1.70 fm.
}
    \label{ba}
\end{figure}

We graph the ratio of the coefficients $B/A$ appearing in eq. 6 in Figure \ref{ba} as a function of  basis size.
For the smaller model spaces this ratio has the wrong sign, but as the model space increases the sign
eventually changes in qualitative agreement with experiment.
 For  $b$ = 1.79 fm, the ratio changes sign between 4$\hbar \omega$ and 6$\hbar \omega$ model spaces;
for the $b$ = 1.94 fm, the sign switches between 8$\hbar \omega$ and 10$\hbar \omega$.

\vspace*{2cm}

\begin{figure}[h]
\includegraphics[width=2.5 in]{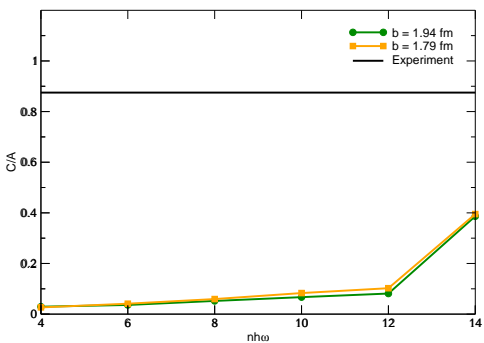}
\caption{(Color online) The ratio $C/A$ (eq. (3)) from a fit to the C2 matrix element in $^6$Li as a function of the
model space.  Experimental data are taken from \cite{stanford,saskatoon,mainz}, and a $C/A$ ratio was extracted using $b$ = 1.70 fm.}
\label{ca}
\end{figure}

Except for the largest model space examined, we see an approximate linear relationship between basis size and $B/A$ and $C/A$, Figs. \ref{ba},\ref{ca}.
The $14\hbar\omega$ model space calculation suggests that the rate of convergence starts to increase faster than this linear relationship suggest.   

\section{Inclusion of a 3-body interaction}
\begin{figure}
\includegraphics[width=2.8in,angle=-90]{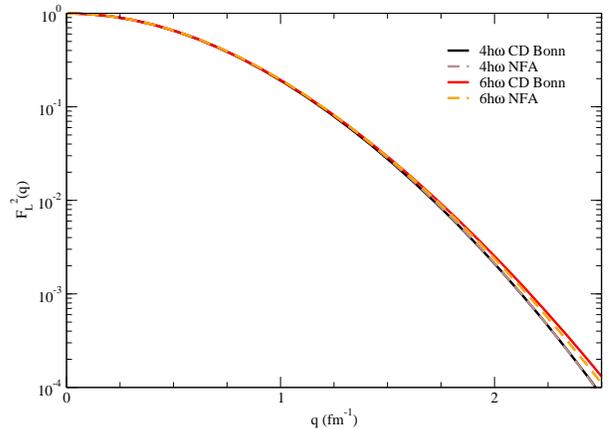}
\caption{(color online) Comparison of the predicted  elastic C0 form factor
for the ground state of $^6$Li with and without the inclusion of a 3-body interaction.}
\label{threebody}
\end{figure}

\begin{figure}
\includegraphics[width=3.0in,]{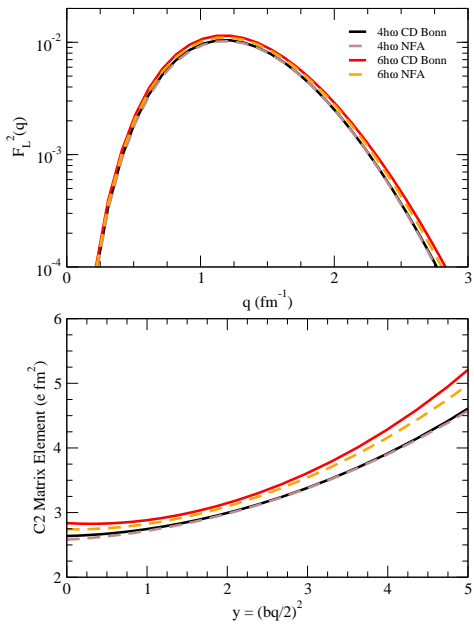}
\caption{ (color online) Comparison of the predicted   C2 form factor
for the  3$^+$ state of $^6$Li with and without the inclusion of a 3-body interaction.}
\label{three2}
\end{figure}

The inclusion of a 3-body interaction leads to an improved predicted level spectrum in $^6$Li, particularly
for the splitting between the ground state and the first 3$^+$ state. In addition, the magnetic form factor for the
$0^+ \rightarrow 1^+$ transition in $^{12}$C is significantly improved when a 3-body interaction is included \cite{navC}. 
 This is because the predicted form factor is very sensitive to the strength of the spin-orbit interaction. However,
 the present $(\Delta L = \lambda$ $ \Delta S=0)$  $C\lambda$ charge form factors  are largely insensitive to
the strength of the spin-orbit interaction and consequently to the 3-body interaction.
In Figure \ref{threebody} we compare the predicted $4\hbar\omega$ and $6\hbar\omega$ predictions for the
elastic C0 form factor for $^{6}$Li. Figure \ref{three2} shows the equivalent calculations for the transition C2
form factor to the 3$^+$ state.
In both cases the inclusion of the 3-body interaction has little effect on the predicted form factor
although it does improve the shape at higher $q$ very slightly.
Figure 7 shows the effect of the 3-body interaction for the inelastic C2 form factor of  $^{12}$C, which
is also very small.

\section{Dependence on the Oscillator Parameter} \label{b-choice}
The unexpected sign for some of the higher shell components in our
 NCSM calculations bears strong resemblance to a similar
problem found in standard multi-$\hbar \omega$ HO shell model calculations.
 When HO standard shell model calculations are extended to include multi-$\hbar\omega$ configurations
 the lack of self-consistency (in the Hartree-Fock sense)
 causes some of the higher shell components in the wave functions to be unphysical \cite{hf1,hf2,hf3,hf4,hf5,hf6,hf7,hf8}.
The main problem arises because  matrix elements of the kinetic energy ($T$) and the two-body interaction ($V$) across two
shells ($\Delta \hbar \omega = 2$) are large and opposite in sign,
 and they cannot be calculated reliably.
The dependence of the C2 matrix element on $b$
implies that the magnitude
 and even the sign of the $\langle ph | T+V | 0\hbar\omega \rangle$ matrix elements depend on $b$.
These off-diagonal matrix elements across two shells in turn  affect
the sign of the leading $ph$ excitations in the wave functions, as well as  all of the similar
$\Delta\hbar\omega = 2$ matrix elements up
to the maximum shell included in the calculation.

\subsection{Dependence of $^6$Li Form factors on the Oscillator Parameter}
\begin{figure}[t]
    \includegraphics[width=3.75 in]{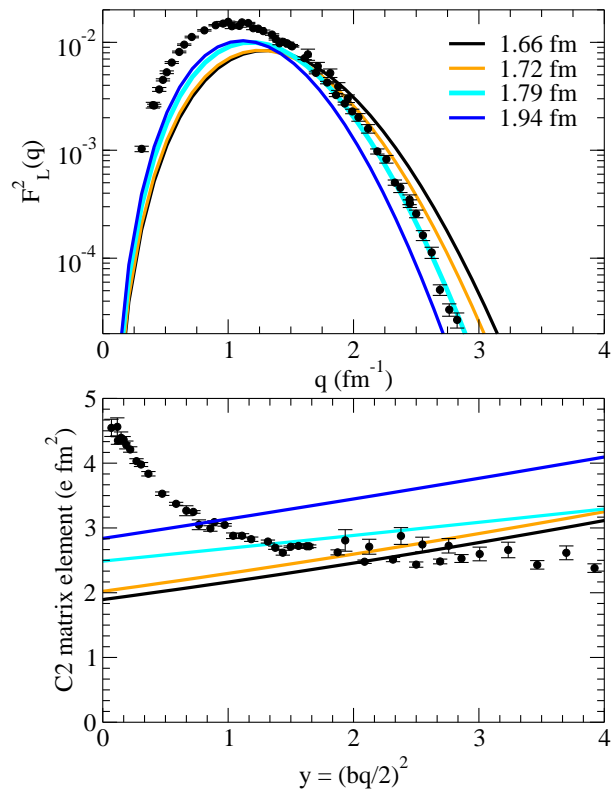}
    \caption{(Color online) The charge form factor and the C2 matrix element for $^6$Li are calculated for several values of $b$ in the 2$\hbar \omega$ model space. }
    \label{both4}
\end{figure}

We investigated the effect of the oscillator parameter on the predicted  form factor within the 2$\hbar \omega$ model space, using
 four different values of $b$ ranging from $b=$ 1.66 - 1.94 fm ($\hbar\omega=$ 15-11 MeV), as shown in
Figure \ref{both4}.  As $b$ varies so does the predicted shape of the form factor: the width of the peak becomes narrower,
and the peak itself  shifts to lower momentum values with increasing $b$.
 The change in slope  of the C2 matrix element with $b$ (Figure \ref{both4}b) suggests  that for sufficiently small $b$,
 the slope of the C2 matrix element will become negative, qualitatively agreeing with experiment.
But such  a small value of $b$ would
would likely result in very slow convergence of the calculations.

\subsection{Dependence of the $^{12}$C Form Factor on the Oscillator Parameter}
\begin{figure}[h]
\includegraphics[width=2.8in]{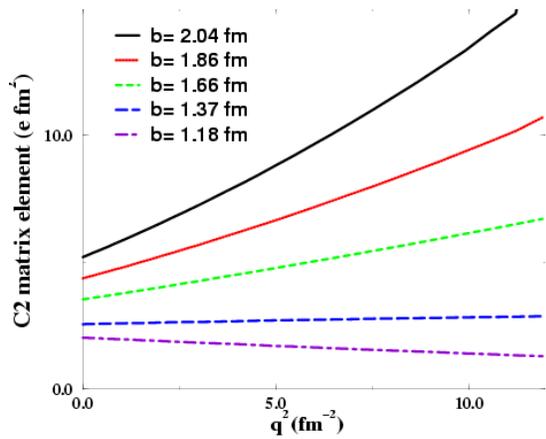}
\caption{(color online) Dependence of the C2 matrix element  for the 4.44 MeV state
in $^{12}$C on the oscillator parameter for the \mbox{$(0+2)\hbar\omega$} model space.}
\label{C2new}
\end{figure}

In Figure \ref{C2new} we display the dependence of the predicted C2 matrix element
in $^{12}$C on the oscillator parameter. These calculations were restricted to a $(0+2)\hbar\omega$ model space.
 As the oscillator parameter is varied, the value  of $\langle ph\mid T+V\mid 0\hbar\omega \rangle$ changes considerably and eventually changes signs.
For sufficiently small $b ( < 1.33 fm)$ the slope of C2($q$) becomes negative, in qualitative agreement with experiment.

\subsection{Dependence of the Giant Resonances on the Oscillator Parameter}
\begin{figure}[h]
\includegraphics[width=2.8in]{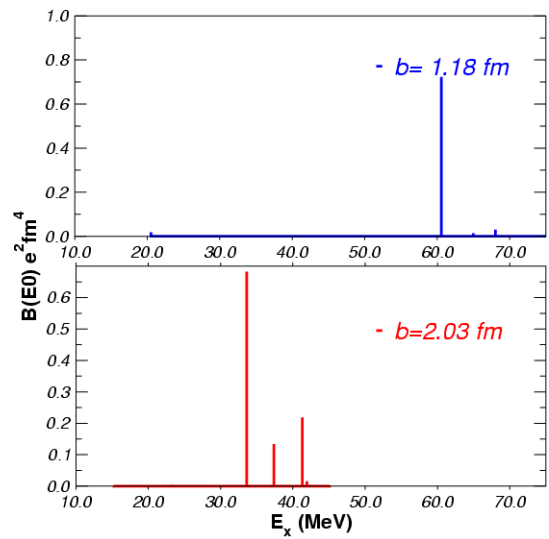}\caption{Dependence of the predicted Giant Monopole Strength (B(E0))
 in $^{12}$C on the oscillator parameter for the $(0+2)\hbar\omega$ model space.}
\label{mono}
\end{figure}

\begin{figure}[h]
\includegraphics[width=2.8in]{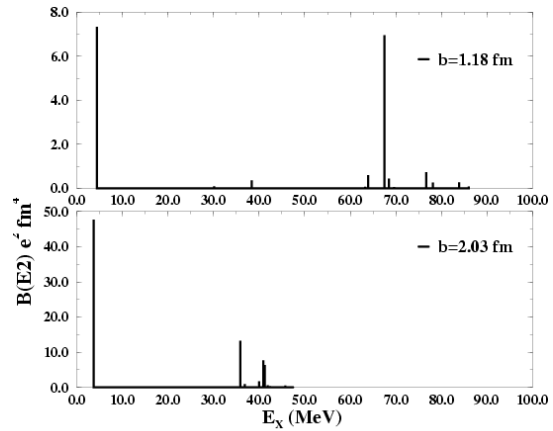}\caption{Dependence of the T=0 E2 strength built on the ground state of
  $^{12}$C on the oscillator parameter for the $(0+2)\hbar\omega$ model space.}
\label{GQR}
\end{figure}

Hoshino et al. \cite{hf7} have pointed out that the problem with the sign of matrix elements of $\langle T+V\rangle$ in multi-$\hbar\omega$
 shell model calculations also manifests itself in the predicted excitation of the GQR and GMR.
The excitation energy of the GMR reflects the compressibility of the nucleus.
Both are intrinsic properties of the nucleon-nucleon interaction;
and, the excitation energy should not depend on the properties of the HO well.
  In Figure \ref{mono} we show the predicted E0 strength for two different values
of the oscillator parameter for our $(0+2)\hbar\omega$ model space.
The large shift in the predicted excitation of the GMR from $\approx 35$ MeV to $\approx 60$ MeV occurs because of the
change in the off-diagonal $\langle ph\mid T + V  \mid 0\hbar\omega \rangle$ matrix elements for the two values of the oscillator parameter.
An analogous problem is seen with the E2 strength, Figure \ref{GQR}.
We note that the sensitivity of the excitation energy of the giant resonances
to the oscillator parameter would likely be  considerably less
for larger model spaces. But our $(0+2)\hbar\omega$ model space calculations exhibit similar sensitivity to
that seen by Hoshino et al. \cite{hf7}. As in the case of the $(e,e')$ form factors, the problem can be directly traced to the $\Delta\hbar\omega=2$, $(\lambda,\mu)=(2,0)$ $ph$ excitations.   


\subsection{Effect of an effective two-body electron scattering operator}
Since the model space sizes discussed here are finite, it is important to address the issue of the impact of effective $(e,e')$
operators and whether they can correct for the sign of the $\Delta\hbar\omega$ = 2 $ph$ excitations in the wave functions.
Stetcu {\it et al.}\cite{stetcu} investigated how a two-body effective contribution affects the E2 and C2 operators.
 In the 0$\hbar \omega$ space, they found that the two-body operator moves the form factor in
the same direction as the larger (0+2)$\hbar \omega$ calculations move; that is, the two-body operator results in the same
unphysical enhancement of
the form factor at large $q$.

\section{Conclusion}
 We have calculated the elastic C0 and the first excited state C2 charge form factors and the corresponding C2 matrix elements in $^6$Li and $^{12}$C within the NCSM using one-body bare operators.  These calculations reveal two primary findings.  First, the magnitude and sign of higher shell $ph$ amplitudes in the wavefunction do not behave as expected.  Higher shell contributions add destructively at low $q$ and constructively at high $q$ to the form factors, contrary to experimental and theoretical expectations.  The relative sign of the symplectic $(\lambda,\mu)=(2,0)$, $\Delta\hbar\omega$ = 2 amplitudes cause them to add destructively to the charge radii.  The large 0$s \rightarrow ns$ (and 0$p \rightarrow np$ for $^{12}$C) amplitudes, introduced by the Lee-Suzuki transformation, also affect the shape of the form factors, further increasing the magnitude of the form factors at high momentum transfers.  In the larger model spaces we explored for $^6$Li, the sign of the 2$\hbar\omega$  contributions to the inelastic form factor changes, but convergence onto experiment is slow.

The second main finding is the strong dependence of the magnitude and sign of the off diagonal $\Delta\hbar\omega$ = 2 matrix elements of $T + V$ on the oscillator parameter.  As a result, the observables (the C2 form factor, GMR, GQR) also heavily depend on the choice of oscillator parameter.  Furthermore, the $ph$ configurations in the low lying wave functions appear with an unexpected sign.  These results  indicates a lack of self-consistency in the NCSM similar to that found in the standard HO shell model.  While there is no known solution, the effects may be minimized by including a Hartree-Fock condition in the calculations   A prescription \cite{hf3, hf7} used in HO shell model calculations is to invoke the following condition:
\begin{equation}
\langle(n+2)\hbar\omega\;\;  ph\mid T+ V\mid n\hbar\omega \rangle^{(\lambda,\mu)=(2,0)}  = 0
\end{equation}
Certainly this method bears further investigation. But in any case a correct treatment  of the symplectic terms in the wave functions is crucial to obtaining a realistic description of electron scattering form factors within a HO shell model basis \cite{Jutta}.  Invoking eq. (7) and/or another prescription to handle the $\Delta\hbar\omega =2$ and higher $ph$ excitations may lead the NCSM to have as much success in predicting momentum-based observables as in predicting energy spectra.

\begin{center}
{\bf \normalsize{Acknowledgments}}
\end{center}
We wish to thank Petr Navr\'atil for providing us with the NCSM one-body
density matrix elements used in these studies, and for many detailed and valuable discussions.
We also wish to thank Ingo Sick, Gerry Peterson and Hall Crannell for providing us with their unpublished electron scattering data.


\begin{thebibliography}{10}
\expandafter\ifx\csname bibnamefont\endcsname\relax
  \def\bibnamefont#1{#1}\fi
\expandafter\ifx\csname bibfnamefont\endcsname\relax
  \def\bibfnamefont#1{#1}\fi
\expandafter\ifx\csname url\endcsname\relax
  \def\url#1{\texttt{#1}}\fi
\expandafter\ifx\csname urlprefix\endcsname\relax\def\urlprefix{URL }\fi
\providecommand{\bibinfo}[2]{#2}
\providecommand{\eprint}[2][]{\url{#2}}

\bibitem{navLi}
\bibinfo{author}{\bibfnamefont{P. Navr\'atil, J.P. Vary, W.E. Ormand, and B.R. Barrett}},
  \bibinfo{journal}{Phys. Rev. Lett.} \textbf{\bibinfo{volume}{87}},
  \bibinfo{pages}{172502} (\bibinfo{year}{2001}).
\bibitem{navEnergy1} P. Navr\'atil, J.P. Vary, B.R. Barrett, Phys. Rev. Lett. {\bf 84}, 5728 (2000).
\bibitem{navEnergy2} P. Navr\'atil, J.P. Vary, W.E. Ormand, and B.R. Barrett, Phys. Rev. C {\bf 62}, 054311 (2000).

\bibitem{forssen}C. Forssen, E. Caurier, P. Navratil, arXiv:0901.0453v1 [nucl-th], (2009).
\bibitem{lee-suzuki} S.Y. Lee and K. Suzuki, Phys. Lett. B {\bf 91} (1980) 173;
\\
K. Suzuki and S.Y. Lee, Prog. Theor. Phys. {\bf 64} (1980) 2091.
\bibitem{navC} A.C. Hayes, P. Navr\'atil, and J.P. Vary, Phys. Rev. Lett. {\bf 99}, 012502 (2003).
\bibitem{dytrych} Tomas Dytrych, Kristina C Sviratcheva, Chairul Bahri, Jerry P. Draayer, James P. Vary,
Phys. Rev. C {\bf 98}, 162503 (2007).
\bibitem{carlson}J. Carlson and R. Schiavilla, Rev. Mod. Phys. {\bf 70}, 743 (1998).
\bibitem{his1} J.C. Bergstrom, Nucl. Phys. {\bf A327}, 458 (1979).
\bibitem{his2} R.M. Hutcheon and H.S. Caplan, Nucl. Phys. {\bf A127}, 417 (1969).
\bibitem{his3} R. Yen, L.S. Cardman, D. Kalinsky, J.R. Legg, and C.K. Bockelman, Nucl. Phys. {\bf A235}, 135 (1974).

\bibitem{stanford} G.C. Li, I. Sick, R.R. Whitney, and M.R. Yearian, Nucl. Phys. {\bf A162}, 583 (1971).
\bibitem{saskatoon} J.C. Bergstrom and E.L. Tomusiak, Nucl. Phys. {\bf A262}, 196 (1976).
\bibitem{mainz} J.C. Bergstrom, U. Deutschmann, R. Neuhausen, Nucl. Phys. {\bf A327}, 439 (1979).
\bibitem{cdbonn} R. Machleidt, F. Sammarruca, and Y. Song, Phys. Rev. C {\bf 53}, 1483 (1996).
\bibitem{joe} R.B. Wiringa and R. Schiavilla, Phys. Rev. Lett. {\bf 81}, 4317 (1998).
\bibitem{exp1}H. L. Crannell and T.A. Griffy, Phys. Rev. {\bf 136} (1964) B1580
\bibitem{exp2}Hall Crannell, Phys. Rev. {\bf 148} (1966) 1107
\bibitem{exp3}A. Nakada, Y. Torizuka, Y. Horikawa, Phys. Rev. Lett. {\bf 27} (1971) 745
\bibitem{exp4}Ingo Sick, {\it private commutation}
\bibitem{millener} D.J. Millener, D.I. Sober, H. Crannell, J.T. O'Brien, L.W. Fagg,
S. Kowalski, C.F. Williams, and L. Lapikas, Phys. Rev. C {\bf 39}, 14 (1989).

\bibitem{Simon}G.G. Simon, Ch. Schmitt, F. Borkowski,, and V.H. Walther, Nucl. Phys. {\bf A333} (1980) 381
\bibitem{tassie} L.J. Tassie and F.C. Barker, Phys. Rev. {\bf 111}, 940 (1958).

\bibitem{AV8}
\bibinfo{author}{\bibfnamefont{B. S. Pudliner {\it et al.}}}
  \bibinfo{journal}{Phys. Rev. C} \textbf{\bibinfo{volume}{56}},
  \bibinfo{pages}{1720} (\bibinfo{year}{1997}).

\bibitem{TM}
\bibinfo{author}{\bibnamefont{{S. A. Coon}}} \bibnamefont{and}
  \bibinfo{author}{\bibnamefont{{H. K. Han}}}, \bibinfo{journal}{Few-Body
  Systems} \textbf{\bibinfo{volume}{30}}, \bibinfo{pages}{131}
  (\bibinfo{year}{2001}).




\bibitem{hf1}S.S.M. Wong, Phys. Lett. {\bf 20}, 188 (1966).
\bibitem{hf2}M.W. Kirson, Nucl. Phys. {\bf A257}, 58 (1976).
\bibitem{hf3} A.C. Hayes, J.L. Friar. and D. Strottman, Phys. Rev. {\bf C 41}, 1727 (1990).
\bibitem{hf4}E.K. Warburton, J.A. Becker, and B.A Brown, Phys. Rev. {\bf C41 }, 1147 (1990).
\bibitem{hf5}W.C. Haxton and C. Johnson, Phys. Rev. Lett. {\bf 65}, 1325 (1990).
\bibitem{hf6}J.P. Blaizot, Phys. Rep. {\bf 64}, 1 (1980).
\bibitem{hf7} Tsutomu Hoshino, Hiroyuki Sagawa and Akito Arima, Nucl. Phys. {\bf A481}, 458 (1988).
\bibitem{hf8}D.J. Millener, A.C. Hayes, and D. Strottman, Phys. Rev {\bf C 45}, 473 (1992).
\bibitem{stetcu}Ionel Stetcu, Bruce R. Barrett, Petr Navratil, and James P. Vary,
Phys. Rev. C. {\bf 71} 044325 (2005), \\
and Ionel Stetcu {\it private communication}.
\bibitem{Jutta}  
Jutta Escher, J. Phys. G {\bf 25} 783-786, (1999).

\end{thebibliography}
\end{document}